\documentclass[english,conference,final]{IEEEtran}
\usepackage[T1]{fontenc}
\usepackage[latin9]{inputenc}
\usepackage{geometry}
\usepackage{amsthm}
\usepackage{amsmath}
\usepackage{amssymb}
\usepackage{esint}

\makeatletter
\theoremstyle{plain}
\newtheorem{thm}{\protect\theoremname}
  \theoremstyle{plain}
  \newtheorem{lem}{\protect\lemmaname}

\usepackage{flushend}
\usepackage[noadjust]{cite}
\usepackage{enumerate}
\usepackage{amsmath}
\usepackage{graphicx}
\usepackage{amssymb}
\usepackage{subfigure}
\usepackage{booktabs}
\usepackage{amsfonts}
\usepackage{amssymb}
\usepackage{url}

\DeclareMathOperator*{\argmax}{arg\,max}

\newcommand{\openone}{\leavevmode\hbox{\small1\normalsize\kern-.33em1}}

\makeatother

\usepackage{babel}
  \providecommand{\lemmaname}{Lemma}
\providecommand{\theoremname}{Theorem}

\begin{document}
\author{
\authorblockN{Jonathan Scarlett} \authorblockA{University of Cambridge\\ \tt{jms265@cam.ac.uk}} 
\and 
\authorblockN{Alfonso Martinez} \authorblockA{Universitat Pompeu Fabra\\ \tt{alfonso.martinez@ieee.org}} 
\and 
\authorblockN{Albert Guill{\'e}n i F{\`a}bregas} \authorblockA{ ICREA \& Universitat Pompeu Fabra\\ University of Cambridge\\ \tt{guillen@ieee.org}}}

\title{A Derivation of the Asymptotic \\
Random-Coding Prefactor}

\maketitle
\long\def\symbolfootnote[#1]#2{\begingroup\def\thefootnote{\fnsymbol{footnote}}\footnote[#1]{#2}\endgroup}
\begin{abstract}
This paper studies the subexponential prefactor to the random-coding
bound for a given rate. Using a refinement of Gallager's bounding
techniques, an alternative proof of a recent result by Altu\u{g}
and Wagner is given, and the result is extended to the setting of
mismatched decoding.
\end{abstract}
\symbolfootnote[0]{This work has been funded in part by the European Research Council under ERC grant agreement 259663, by the European Union's 7th Framework Programme (PEOPLE-2011-CIG) under grant agreement 303633 and by the Spanish Ministry of Economy and Competitiveness under grants RYC-2011-08150 and TEC2012-38800-C03-03.}

\section{Introduction \label{sec:INTRO}}

Error exponents are a widely-studied tool in information theory for
characterizing the performance of coded communication systems. Early
works on error exponents for discrete memoryless channels (DMCs) include
those of Fano \cite[Ch. 9]{FanoBook}, Gallager \cite[Ch. 5]{Gallager}
and Shannon \emph{et al. }\cite{SpherePacking1}. The achievable exponent
of \cite{FanoBook,Gallager} was obtained using i.i.d. random coding,
and coincides with the sphere-packing exponent given in \cite{SpherePacking1}
for rates above a threshold called the critical rate.

Denoting the exponent of \cite{FanoBook,Gallager} by $E_{r}(R)$,
we have the following: For all $(n,R)$, there exists a code of rate
$R$ and block length $n$ such that the error probability $p_{e}$
satisfies $p_{e}\le\alpha(n,R)e^{-nE_{r}(R)}$, where $\alpha(n,R)$
is a subexponential prefactor. In both \cite{FanoBook} and \cite{Gallager},
the prefactor is $O(1)$. In particular, Gallager showed that one
can achieve $\alpha(n,R)=1$.

Early works on improving the $O(1)$ prefactor for certain channels
and rates include those of Elias \cite{TwoChannels}, Dobrushin \cite{Dobrushin}
and Gallager \cite{TightAverage}. These results were recently generalized
by Altu\u{g}  and Wagner \cite{RefinementRC,RefinementRC2,RefinementSP},
who obtained prefactors to the random-coding bound at all rates below
capacity, as well as converse results above the critical rate. The
bounds in \cite{RefinementRC,RefinementRC2} were obtained using i.i.d.
random coding, and the behavior of the prefactor varies depending
on whether the rate is above or below the critical rate, and whether
a regularity condition is satisfied (see Section\textbf{ }\ref{sec:REF_MAIN_RESULT}).

In this paper, we give an alternative proof of the main result of
\cite{RefinementRC,RefinementRC2}, as well as a generalization to
the setting of mismatched decoding \cite{Csiszar2,Merhav,Compound,MMRevisited,JournalSU},
where the decoding rule is fixed and possibly suboptimal (e.g. due
to channel uncertainty or implementation constraints). The analysis
of \cite{RefinementRC,RefinementRC2} can be considered a refinement
of that of Fano \cite[Ch. 9]{FanoBook}, whereas the analysis in this
paper can be considered a refinement of that of Gallager \cite[Ch. 5]{Gallager}.
Our techniques can also be used to derive Gallager's expurgated exponent
\cite[Ch. 5.7]{Gallager} with an $O\big(\frac{1}{\sqrt{n}}\big)$
prefactor under some technical conditions \cite{PaperExpurg}, thus
improving on Gallager's $O(1)$ prefactor.

\subsection{Notation }

Vectors are written using bold symbols (e.g. $\boldsymbol{x}$), and
the corresponding $i$-th entry is written with a subscript (e.g.
$x_{i}$). For two sequences $f_{n}$ and $g_{n}$, we write $f_{n}=O(g_{n})$
if $|f_{n}|\le c|g_{n}|$ for some $c$ and sufficiently large $n$,
and $f_{n}=o(g_{n})$ if $\lim_{n\to\infty}\frac{f_{n}}{g_{n}}=0$.
The indicator function is denoted by $\openone\{\cdot\}$.

The marginals of a joint  distribution $P_{XY}(x,y)$ are denoted
by $P_{X}(x)$ and $P_{Y}(y)$. Expectation with respect to a joint
distribution $P_{XY}(x,y)$ is denoted by $\mathbb{E}_{P}[\cdot]$,
or simply $\mathbb{E}[\cdot]$ when the probability distribution is
understood from the context. Given a distribution $Q(x)$ and conditional
distribution $W(y|x)$, we write $Q\times W$ to denote the joint
distribution defined by $Q(x)W(y|x)$. The set of all empirical distributions
on a vector in $\mathcal{X}^{n}$ (i.e. types \cite[Sec. 2]{CsiszarBook},
\cite{GallagerCC}) is denoted by $\mathcal{P}_{n}(\mathcal{X})$.
The type of a vector $\boldsymbol{x}$ is denoted by $\hat{P}_{\boldsymbol{x}}(\cdot)$.
For a given $Q\in\mathcal{P}_{n}(\mathcal{X})$, the type class $T^{n}(Q)$
is defined to be the set of sequences in $\mathcal{X}^{n}$ with type
$Q$.

\section{Statement of Main Result \label{sec:REF_MAIN_RESULT}}

Let $\mathcal{X}$ and $\mathcal{Y}$ denote the input and output
alphabets respectively. The probability of receiving a given output
sequence $\boldsymbol{y}$ given that $\boldsymbol{x}$ is transmitted
is given by $W^{n}(\boldsymbol{y}|\boldsymbol{x})\stackrel{\triangle}{=}\prod_{i=1}^{n}W(y_{i}|x_{i})$.
A codebook $\mathcal{C}=\{\boldsymbol{x}^{(1)},...,\boldsymbol{x}^{(M)}\}$
is known at both the encoder and decoder. The encoder receives as
input a message $m$ uniformly distributed on the set $\{1,...,M\}$,
and transmits the corresponding codeword $\boldsymbol{x}^{(m)}$.
Given $\boldsymbol{y}$, the decoder forms the estimate 
\begin{equation}
\hat{m}=\argmax_{j\in\{1,...,M\}}q^{n}(\boldsymbol{x}^{(j)},\boldsymbol{y}),\label{eq:REF_DecodingRule}
\end{equation}
where $n$ is the block length, and $q^{n}(\boldsymbol{x},\boldsymbol{y})\triangleq\prod_{i=1}^{n}q(x_{i},y_{i})$.
The function $q(x,y)$ is called the \emph{decoding metric}, and is
assumed to be non-negative and such that
\begin{equation}
q(x,y)=0\iff W(y|x)=0.\label{eq:REF_Assumption1}
\end{equation}
In the case of a tie, a random codeword achieving the maximum in (\ref{eq:REF_DecodingRule})
is selected. In the case that $q(x,y)=W(y|x)$, i.e. maximum-likelihood
(ML) decoding, the decoding rule in (\ref{eq:REF_DecodingRule}) is
optimal. Otherwise, this setting is that of \emph{mismatched decoding}
\cite{Csiszar2,Merhav,Compound,MMRevisited,JournalSU}. 

We study the random-coding error probability under i.i.d. random coding,
where the $M=e^{nR}$ codewords are generated independently according
to
\begin{equation}
P_{\boldsymbol{X}}(\boldsymbol{x})=Q^{n}(\boldsymbol{x})\stackrel{\triangle}{=}\prod_{i=1}^{n}Q(x_{i}),\label{eq:REF_Px_IID}
\end{equation}
and where $Q$ is an arbitrary input distribution. The random-coding
error probability is denoted by $\overline{p}_{e}$. 

\addtolength{\topmargin}{-0.25in}

It was shown in \cite{Compound} that $\overline{p}_{e}\le e^{-nE_{r}(Q,R)}$,
where
\begin{align}
E_{r}(Q,R) & \stackrel{\triangle}{=}\max_{\rho\in[0,1]}E_{0}(Q,\rho)-\rho R\label{eq:REF_Er_IID}\\
E_{0}(Q,\rho) & \stackrel{\triangle}{=}\sup_{s\ge0}-\log\mathbb{E}\left[\bigg(\frac{\mathbb{E}\big[q(\overline{X},Y)^{s}\,|\, Y\big]}{q(X,Y)^{s}}\bigg)^{\rho}\right]\label{eq:REF_E0_IID}
\end{align}
with $(X,Y,\overline{X})\sim Q(x)W(y|x)Q(\overline{x})$. We showed
in \cite{PaperSU} that this exponent is tight with respect to the
ensemble average for i.i.d. random coding, i.e. $\lim_{n\to\infty}-\frac{1}{n}\log\overline{p}_{e}=E_{r}$.
The corresponding achievable rate is given by 
\begin{equation}
I_{\mathrm{GMI}}(Q)\stackrel{\triangle}{=}\sup_{s\ge0}\mathbb{E}\bigg[\log\frac{q(X,Y)^{s}}{\mathbb{E}\big[q(\overline{X},Y)^{s}\,|\, Y\big]}\bigg],\label{eq:REF_GMI}
\end{equation}
which is commonly referred to as the generalized mutual information
(GMI) \cite{Compound}. Under ML decoding, i.e. $q(x,y)=W(y|x)$,
$E_{r}$ equals the exponent of Fano and Gallager \cite{FanoBook,Gallager},
and $I_{\mathrm{GMI}}(Q)$ equals the mutual information. The corresponding
optimal choices of $s$ in (\ref{eq:REF_E0_IID})--(\ref{eq:REF_GMI})
are respectively given by $s=\frac{1}{1+\rho}$ and $s=1$.

We define $\hat{\rho}(Q,R)$ to be the value of $\rho$ achieving
the maximum in (\ref{eq:REF_Er_IID}) at rate $R$. From the analysis
of Gallager \cite[Sec. 5.6]{Gallager}, we know that $\hat{\rho}$
equals one for all rates between $0$ and some critical rate,
\begin{equation}
R_{\mathrm{cr}}(Q)\stackrel{\triangle}{=}\max\big\{ R\,:\,\hat{\rho}(Q,R)=1\big\},
\end{equation}
and is strictly decreasing for all rates between $R_{\mathrm{cr}}(Q)$
and $I_{\mathrm{GMI}}(Q)$.

Similarly to \cite{RefinementRC}, we define the following notion
of regularity. We introduce the set
\begin{multline}
\mathcal{Y}_{1}\stackrel{\triangle}{=}\Big\{ y\,:\, q(x,y)\ne q(\overline{x},y)\text{ for some }\\
x,\overline{x}\text{ such that }Q(x)Q(\overline{x})W(y|x)W(y|\overline{x})>0\Big\}\label{eq:REF_SetY1}
\end{multline}
and define $(W,q,Q)$ to be \emph{regular }if
\begin{equation}
\mathcal{Y}_{1}\ne\emptyset.\label{eq:REF_Assumption2}
\end{equation}
When $q(x,y)=W(y|x)$, this is the \emph{feasibility decoding is suboptimal
}(FDIS) condition of \cite{RefinementRC}. We say that $(W,q,Q)$
is \emph{irregular }if it is not regular. A notable example of the
irregular case is the binary erasure channel (BEC) under ML decoding.
\begin{thm}
\label{thm:REF_Refinement}Fix any $(W,q)$ satisfying (\ref{eq:REF_Assumption1}),
input distribution $Q$ and rate $R<I_{\mathrm{GMI}}(Q)$. The random-coding
error probability for the i.i.d. ensemble in (\ref{eq:REF_Px_IID})
satisfies 
\begin{equation}
\overline{p}_{e}\le\alpha(n,R)e^{-nE_{r}(Q,R)}
\end{equation}
for sufficiently large $n$, where $\alpha(n,R)$ is defined as follows.
If $(W,q,Q)$ is regular, then 
\begin{equation}
\alpha(n,R)\stackrel{\triangle}{=}\begin{cases}
\frac{K}{n^{\frac{1}{2}(1+\hat{\rho}(Q,R))}} & R\in\big(R_{\mathrm{cr}}(Q),I_{\mathrm{GMI}}(Q)\big)\\
\frac{K}{\sqrt{n}} & R\in\big[0,R_{\mathrm{cr}}(Q)\big],
\end{cases}\label{eq:REF_alpha}
\end{equation}
and if $(W,q,Q)$ is irregular, then 
\begin{equation}
\alpha(n,R)\stackrel{\triangle}{=}\begin{cases}
\frac{K}{\sqrt{n}} & R\in\big(R_{\mathrm{cr}}(Q),I_{\mathrm{GMI}}(Q)\big)\\
1 & R\in\big[0,R_{\mathrm{cr}}(Q)\big],
\end{cases}\label{eq:REF_alpha-1}
\end{equation}
where $K$ is a constant depending only on $W$, $q$, $Q$ and $R$.\end{thm}
\begin{IEEEproof}
See Section \ref{sec:REF_PROOF}.
\end{IEEEproof}
In the case of ML decoding, Theorem \ref{thm:REF_Refinement} coincides
with the main results of Altu\u{g}  and Wagner \cite{RefinementRC,RefinementRC2}
in both the regular and irregular case. Neither \cite{RefinementRC,RefinementRC2}
nor the present paper attempt to explicitly characterize or bound
the constant $K$ in (\ref{eq:REF_alpha})--(\ref{eq:REF_alpha-1}).
Asymptotic bounds with the constant factor specified are derived in
\cite{JournalSU} using saddlepoint approximations; see also \cite{TightAverage}
for rates below the critical rate, and \cite{Dobrushin} for strongly
symmetric channels.%
\footnote{The English translation of \cite{Dobrushin} incorrectly states that
the prefactor is $O\big(n^{-\frac{1}{2(1+\hat{\rho}(R))}}\big)$ for
the regular case with $R>R_{\mathrm{cr}}$ (see (1.28)--(1.32) therein),
but this error is not present in the original Russian version.%
}

\section{\label{sec:REF_PROOF}Proof of Theorem \ref{thm:REF_Refinement}}

For a fixed value of $s\ge0$, we define the \emph{generalized information
density }\cite{Finite,PaperSU}
\begin{align}
i_{s}(x,y) & \stackrel{\triangle}{=}\log\frac{q(x,y)^{s}}{\sum_{\overline{x}}Q(\overline{x})q(\overline{x},y)^{s}}\label{eq:REF_is}
\end{align}
and its multi-letter extension
\begin{equation}
i_{s}^{n}(\boldsymbol{x},\boldsymbol{y})\stackrel{\triangle}{=}\sum_{i=1}^{n}i_{s}(x_{i},y_{i}).\label{eq:REF_is_n}
\end{equation}
Our analysis is based on the random-coding union (RCU) bound for mismatched
decoding, given by \cite{Finite,PaperSU}
\begin{multline}
\overline{p}_{e}\le\mathbb{E}\bigg[\min\Big\{1,(M-1)\\
\times\mathbb{P}\big[i_{s}^{n}(\overline{\boldsymbol{X}},\boldsymbol{Y})\ge i_{s}^{n}(\boldsymbol{X},\boldsymbol{Y})\,|\,\boldsymbol{X},\boldsymbol{Y}\big]\Big\}\bigg],\label{eq:REF_RCU}
\end{multline}
where $(\boldsymbol{X},\boldsymbol{Y},\overline{\boldsymbol{X}})\sim P_{\boldsymbol{X}}(\boldsymbol{x})W^{n}(\boldsymbol{y}|\boldsymbol{x})P_{\boldsymbol{X}}(\overline{\boldsymbol{x}})$.
Furthermore, we will make use of the identity
\begin{equation}
E_{0}(Q,\rho)=\sup_{s\ge0}-\log\mathbb{E}\big[e^{-\rho i_{s}(X,Y)}\big]\label{eq:REF_E0_is}
\end{equation}
with $(X,Y)\sim Q\times W$, which follows from (\ref{eq:REF_E0_IID})
and (\ref{eq:REF_is}).

We provide a number of preliminary results in Section \ref{sub:REF_PRELIM}.
The proof of Theorem \ref{thm:REF_Refinement} for the regular case
is given in Section \ref{sub:REF_PROOF_REGULAR}, and the changes
required to handle the irregular case are given in Section \ref{sub:REF_PROOF_IRREGULAR}.

\subsection{\label{sub:REF_PRELIM}Preliminary Results}

The main tool used in the proof of Theorem \ref{thm:REF_Refinement}
is the following lemma by Polyanskiy \emph{et al. }\cite{Finite},
which can be proved using the Berry-Esseen theorem.
\begin{lem}
\emph{\label{lem:REF_Lem20}\cite[Lemma 47]{Finite}~}Let $Z_{1},...,Z_{n}$
be independent random variables with $\sigma^{2}=\sum_{i=1}^{n}\mathrm{Var}[Z_{i}]>0$
and $T=\sum_{i=1}^{n}\mathbb{E}[|Z_{i}-\mathbb{E}[Z_{i}]|^{3}]<\infty$.
Then for any $t$,\emph{
\begin{multline}
\mathbb{E}\bigg[\exp\Big(-\sum_{i}Z_{i}\Big)\openone\Big\{\sum_{i}Z_{i}>t\Big\}\bigg]\\
\le2\Big(\frac{\log2}{\sqrt{2\pi}}+\frac{12T}{\sigma^{2}}\Big)\frac{1}{\sigma}\exp\big(-t\big).\label{eq:REF_Lem20}
\end{multline}
}
\end{lem}
The following lemma shows that under the assumption (\ref{eq:REF_Assumption1}),
we do not need to consider $s$ growing unbounded in (\ref{eq:REF_E0_IID}).
\begin{lem}
\label{lem:REF_sLem}For any $(W,q)$ satisfying (\ref{eq:REF_Assumption1}),
and any $\rho\in[0,1]$, the supremum in (\ref{eq:REF_E0_IID}) is
achieved (possibly non-uniquely) by some finite $s\ge0$.\end{lem}
\begin{IEEEproof}
We treat the regular and irregular cases separately. In the regular
case, let $(x,\overline{x},y)$ satisfy the condition in the definition
of $\mathcal{Y}_{1}$ in (\ref{eq:REF_SetY1}), and assume without
loss of generality that $q(\overline{x},y)>q(x,y)$. We can upper
bound the objective in (\ref{eq:REF_E0_IID}) by 
\begin{equation}
-\log Q(x)W(y|x)\bigg(Q(\overline{x})\bigg(\frac{q(\overline{x},y)}{q(x,y)}\bigg)^{s}\bigg)^{\rho},\label{eq:REF_sLem1}
\end{equation}
which tends to $-\infty$ as $s\to\infty$. It follows that the supremum
is achieved by a finite value of $s$. 

In the irregular case, we have $q(x,y)=q(\overline{x},y)$ wherever
$Q(x)Q(\overline{x})W(y|x)q(\overline{x},y)>0$, where the replacement
of $W(y|\overline{x})$ by $q(\overline{x},y)$ in the latter condition
follows from (\ref{eq:REF_Assumption1}). In this case, writing the
objective in (\ref{eq:REF_E0_IID}) as
\begin{equation}
-\log\sum_{x,y}Q(x)W(y|x)\bigg(\sum_{\overline{x}}Q(\overline{x})\bigg(\frac{q(\overline{x},y)}{q(x,y)}\bigg)^{s}\bigg)^{\rho},
\end{equation}
we see that all choices of $s>0$ are equivalent, since the argument
to $(\cdot)^{s}$ equals one for all $(x,\overline{x},y)$ yielding
non-zero terms in the summations.
\end{IEEEproof}
The following lemma is somewhat more technical, and ensures the existence
of a sufficiently high probability set in which Lemma \ref{lem:REF_Lem20}
can be applied to the inner probability in (\ref{eq:REF_ExpandedRCU})
with a value of $\sigma$ having $\sqrt{n}$ growth. We make use of
the conditional distributions
\begin{align}
V_{s}(x|y) & \stackrel{\triangle}{=}\frac{Q(x)q(x,y)^{s}}{\sum_{\overline{x}}Q(\overline{x})q(\overline{x},y)^{s}}\label{eq:REF_Vs}\\
V_{s}^{n}(\boldsymbol{x}|\boldsymbol{y}) & \stackrel{\triangle}{=}\prod_{i=1}^{n}V_{s}(x_{i}|y_{i}),
\end{align}
which yield $i_{s}(x,y)=\log\frac{V_{s}(x|y)}{Q(x)}$ and $i_{s}^{n}(\boldsymbol{x},\boldsymbol{y})=\log\frac{V_{s}^{n}(\boldsymbol{x}|\boldsymbol{y})}{Q^{n}(\boldsymbol{x})}$
(see (\ref{eq:REF_is})--(\ref{eq:REF_is_n})). Furthermore, we define
the random variables
\begin{align}
(X,Y,\overline{X},X_{s}) & \sim Q(x)W(y|x)Q(\overline{x})V_{s}(x_{s}|y)\nonumber \\
(\boldsymbol{X},\boldsymbol{Y},\overline{\boldsymbol{X}},\boldsymbol{X}_{s}) & \sim Q^{n}(\boldsymbol{x})W^{n}(\boldsymbol{y}|\boldsymbol{x})Q^{n}(\overline{\boldsymbol{x}})V_{s}^{n}(\boldsymbol{x}_{s}|\boldsymbol{y}),\label{eq:REF_BoldVars}
\end{align}
and we write the empirical distribution of $\boldsymbol{y}$ as $\hat{P}_{\boldsymbol{y}}(\cdot)$.
\begin{lem}
\emph{\label{lem:SetF}}If $(W,q,Q)$\emph{ is regular and }(\ref{eq:REF_Assumption1})\emph{
holds, then the set
\begin{equation}
\mathcal{F}_{n,\delta}\stackrel{\triangle}{=}\Big\{\boldsymbol{y}\,:\,\sum_{y\in\mathcal{Y}_{1}}\hat{P}_{\boldsymbol{y}}(y)>\delta\Big\}\label{eq:SetFn}
\end{equation}
satisfies the following properties: }\end{lem}
\begin{enumerate}
\item \emph{For any $\boldsymbol{y}\in\mathcal{F}_{n,\delta}$, we have
\begin{equation}
\mathrm{Var}\big[i_{s}^{n}(\boldsymbol{X}_{s},\boldsymbol{Y})\,|\,\boldsymbol{Y}=\boldsymbol{y}\big]\ge n\delta v_{s},\label{eq:VarLB}
\end{equation}
 where
\begin{equation}
v_{s}\stackrel{\triangle}{=}\min_{y\in\mathcal{Y}_{1}}\mathrm{Var}\big[i_{s}(X_{s},Y)\,|\, Y=y\big].\label{eq:v_s}
\end{equation}
Furthermore, $v_{s}>0$ for all $s>0$.}
\item \emph{For all }$R<I_{\mathrm{GMI}}(Q)$\emph{, there exists a choice
of $\delta>0$ such that under i.i.d. random coding,
\begin{equation}
\mathbb{P}\big[\mathrm{error}\,\cap\,\boldsymbol{Y}\notin\mathcal{F}_{n,\delta}\big]\le e^{-n(E_{r}^{\prime}(Q,R)+o(1))}\label{eq:ErrorFn}
\end{equation}
for some $E_{r}^{\prime}(Q,R)>E_{r}(Q,R)$.}\end{enumerate}
\begin{IEEEproof}
See the Appendix.
\end{IEEEproof}

\subsection{\label{sub:REF_PROOF_REGULAR}Proof for the Regular Case}

Using the second part of Lemma \ref{lem:SetF} with the suitably chosen
value of $\delta$, and using the fact that $\lim_{n\to\infty}-\frac{1}{n}\log\overline{p}_{e}=E_{r}$
\cite{PaperSU}, we can write the random-coding error probability
as
\begin{align}
\overline{p}_{e} & =\mathbb{P}\big[\mathrm{error}\,\cap\,\boldsymbol{Y}\in\mathcal{F}_{n,\delta}\big]+\mathbb{P}\big[\mathrm{error}\,\cap\,\boldsymbol{Y}\notin\mathcal{F}_{n,\delta}\big]\\
 & =\big(1+o(1)\big)\mathbb{P}\big[\mathrm{error}\,\cap\,\boldsymbol{Y}\in\mathcal{F}_{n,\delta}\big].\label{eq:REF_MainProof2}
\end{align}
Writing $K_{1}$ in place of $1+o(1)$ and modifying the RCU bound
in (\ref{eq:REF_RCU}) to include the condition $\boldsymbol{Y}\in\mathcal{F}_{n,\delta}$
in (\ref{eq:REF_MainProof2}), we obtain 
\begin{multline}
\overline{p}_{e}\le K_{1}\sum_{\boldsymbol{x},\boldsymbol{y}\in\mathcal{F}_{n,\delta}}P_{\boldsymbol{X}}(\boldsymbol{x})W^{n}(\boldsymbol{y}|\boldsymbol{x})\\
\times\min\Big\{1,M\mathbb{P}\big[i_{s}^{n}(\overline{\boldsymbol{X}},\boldsymbol{y})\ge i_{s}^{n}(\boldsymbol{x},\boldsymbol{y})\big]\Big\}.\label{eq:REF_ExpandedRCU}
\end{multline}
The value of $s\ge0$ in (\ref{eq:REF_ExpandedRCU}) is arbitrary,
and we choose it to achieve the supremum in (\ref{eq:REF_E0_IID})
at $\rho=\hat{\rho}(Q,R)$, in accordance with Lemma \ref{lem:REF_sLem}.
We can assume that $s>0$, since $s=0$ yields an objective of zero
in (\ref{eq:REF_E0_IID}), contradicting the assumption that $R<I_{\mathrm{GMI}}$. 

In order to make the inner probability in (\ref{eq:REF_ExpandedRCU})
more amenable to an application of Lemma \ref{lem:REF_Lem20}, we
follow \cite[Sec. 3.4.5]{FiniteThesis} and write
\begin{align}
Q^{n}(\overline{\boldsymbol{x}}) & =Q^{n}(\overline{\boldsymbol{x}})\frac{V_{s}^{n}(\overline{\boldsymbol{x}}|\boldsymbol{y})}{V_{s}^{n}(\overline{\boldsymbol{x}}|\boldsymbol{y})}\\
 & =V_{s}^{n}(\overline{\boldsymbol{x}}|\boldsymbol{y})\exp\big(-i_{s}^{n}(\overline{\boldsymbol{x}},\boldsymbol{y})\big).\label{eq:eq:iidDeriv0}
\end{align}
For a fixed sequence $\boldsymbol{y}$ and a constant $t$, summing
both sides of (\ref{eq:eq:iidDeriv0}) over all $\overline{\boldsymbol{x}}$
such that $i_{s}^{n}(\overline{\boldsymbol{x}},\boldsymbol{y})\ge t$
yields
\begin{multline}
\mathbb{P}\big[i_{s}^{n}(\overline{\boldsymbol{X}},\boldsymbol{y})\ge t\big]\\
=\mathbb{E}\Big[\exp\big(-i_{s}^{n}(\boldsymbol{X}_{s},\boldsymbol{Y})\big)\openone\big\{ i_{s}^{n}(\boldsymbol{X}_{s},\boldsymbol{Y})\ge t\big\}\,\Big|\,\boldsymbol{Y}=\boldsymbol{y}\Big]\label{eq:iidDeriv1}
\end{multline}
under the joint distribution in (\ref{eq:REF_BoldVars}). Applying
Lemma \ref{lem:REF_Lem20} to (\ref{eq:iidDeriv1}) and using the
first part of Lemma \ref{lem:SetF}, we obtain for all $\boldsymbol{y}\in\mathcal{F}_{n,\delta}$
that 
\begin{multline}
\mathbb{E}\Big[\exp\big(-i_{s}^{n}(\boldsymbol{X}_{s},\boldsymbol{Y})\big)\openone\big\{ i_{s}^{n}(\boldsymbol{X}_{s},\boldsymbol{Y})\ge t\big\}\,\Big|\,\boldsymbol{Y}=\boldsymbol{y}\Big]\\
\le\frac{K_{2}}{\sqrt{n}}e^{-t}\label{eq:REF_Tail1}
\end{multline}
for some constant $K_{2}$. Here we have used the fact that $T$ in
(\ref{eq:REF_Lem20}) grows linearly in $n$, which follows from the
fact that we are considering finite alphabets \cite[Lemma 46]{Finite}.
Substituting (\ref{eq:REF_Tail1}) into (\ref{eq:REF_ExpandedRCU}),
we obtain 
\begin{align}
\overline{p}_{e} & \le K_{1}\sum_{\boldsymbol{x},\boldsymbol{y}\in\mathcal{F}_{n,\delta}}P_{\boldsymbol{X}}(\boldsymbol{x})W^{n}(\boldsymbol{y}|\boldsymbol{x})\\
 & \qquad\qquad\qquad\times\min\bigg\{1,\frac{MK_{2}}{\sqrt{n}}e^{-i_{s}^{n}(\boldsymbol{x},\boldsymbol{y})}\bigg\}\\
 & \le K_{1}\mathbb{E}\bigg[\min\bigg\{1,\frac{MK_{2}}{\sqrt{n}}e^{-i_{s}^{n}(\boldsymbol{X},\boldsymbol{Y})}\bigg\}\bigg]\label{eq:REF_Step2}\\
 & \le K_{3}\mathbb{E}\bigg[\min\bigg\{1,\frac{M}{\sqrt{n}}e^{-i_{s}^{n}(\boldsymbol{X},\boldsymbol{Y})}\bigg\}\bigg]\label{eq:REF_Step3}
\end{align}
where (\ref{eq:REF_Step2}) follows by upper bounding the summation
over $\boldsymbol{y}\in\mathcal{F}_{n,\delta}$ by a summation over
all $\boldsymbol{y}$, and (\ref{eq:REF_Step3}) follows by defining
$K_{3}\stackrel{\triangle}{=}K_{1}\max\{1,K_{2}\}$. 

We immediately obtain the desired result for rates below the critical
rate by upper bounding the $\min\{1,\cdot\}$ term in (\ref{eq:REF_Step3})
by one and using (\ref{eq:REF_E0_is}) (with $\rho=1$) and the definition
of $i_{s}^{n}$. In the remainder of the subsection, we focus on rates
above the critical rate.

For any non-negative random variable $A$, we have $\mathbb{E}[\min\{1,A\}]=\mathbb{P}[A\ge U]$,
where $U$ is uniform on $(0,1)$ and independent of $A$. We can
thus write (\ref{eq:REF_Step3}) as 
\begin{align}
\overline{p}_{e} & \le K_{3}\mathbb{P}\bigg[\frac{M}{\sqrt{n}}e^{-i_{s}^{n}(\boldsymbol{X},\boldsymbol{Y})}\ge U\bigg]\\
 & =K_{3}\mathbb{P}\bigg[\sum_{i=1}^{n}\big(R-i_{s}(X_{i},Y_{i})\big)\ge\log\big(U\sqrt{n}\big)\bigg].\label{eq:REF_Step5}
\end{align}
Let $F(t)$ denote the cumulative distribution function (CDF) of $R-i_{s}(X,Y)$
with $(X,Y)\sim Q\times W$, and let $Z_{1},\cdots,Z_{n}$ be i.i.d.
according to the tilted CDF 
\begin{equation}
F_{Z}(z)=e^{E_{r}(Q,R)}\int_{-\infty}^{z}e^{\hat{\rho}t}dF(t),\label{eq:SA_Above_2-2}
\end{equation}
where $\hat{\rho}=\hat{\rho}(Q,R)$. It is easily seen that this is
indeed a CDF by writing
\begin{equation}
\int_{-\infty}^{\infty}e^{\hat{\rho}t}dF(t)=\mathbb{E}\big[e^{\hat{\rho}(R-i_{s}(X,Y))}\big]=e^{-E_{r}(Q,R)},
\end{equation}
where the last equality follows from (\ref{eq:REF_E0_is}) and since
we have assumed that $s$ is chosen optimally.

Similarly to \cite[Lemma 2]{BahadurRao}, we can use (\ref{eq:SA_Above_2-2})
to write the probability in (\ref{eq:REF_Step5}) as follows:
\begin{align}
 & \mathbb{P}\bigg[\sum_{i=1}^{n}\big(R-i_{s}(X_{i},Y_{i})\big)\ge\log\big(U\sqrt{n}\big)\bigg]\nonumber \\
 & =\dotsint_{\sum_{i}t_{i}\ge\log(u\sqrt{n})}dF(t_{1})\cdots dF(t_{n})dF_{U}(u)\\
 & =e^{-nE_{r}(Q,R)}\dotsint_{\sum_{i}z_{i}\ge\log(u\sqrt{n})}e^{-\hat{\rho}\sum_{i}z_{i}}\nonumber \\
 & \qquad\qquad\qquad\qquad\times dF_{Z}(z_{1})\cdots dF_{Z}(z_{n})dF_{U}(u),\label{eq:SA_ChgMeasure2}
\end{align}
where $F_{U}(u)$ denotes the CDF of $U$. Substituting (\ref{eq:SA_ChgMeasure2})
into (\ref{eq:REF_Step5}), we obtain 
\begin{multline}
\overline{p}_{e}\le K_{3}e^{-nE_{r}(Q,R)}\\
\times\mathbb{E}\Big[e^{-\hat{\rho}\sum_{i}Z_{i}}\openone\Big\{\hat{\rho}\sum_{i}Z_{i}\ge\hat{\rho}\log\big(U\sqrt{n}\big)\Big\}\Big].\label{eq:SA_Above_3}
\end{multline}

Let $E_{0}(Q,\rho,s)$ be defined as in (\ref{eq:REF_E0_IID}) with
a fixed value of $s$ in place of the supremum. The moment generating
function (MGF) of $Z$ is given by
\begin{align}
M_{Z}(\tau) & =\mathbb{E}[e^{\tau Z}]\\
 & =e^{E_{r}(Q,R)}\mathbb{E}[e^{(\hat{\rho}+\tau)(R-i_{s}(X,Y))}]\label{eq:REF_Mz2}\\
 & =e^{E_{0}(Q,\hat{\rho},s)}e^{-(E_{0}(Q,\hat{\rho}+\tau,s)-\tau R)},\label{eq:REF_Mz3}
\end{align}
where (\ref{eq:REF_Mz2}) follows from (\ref{eq:SA_Above_2-2}), and
(\ref{eq:REF_Mz3}) follows from (\ref{eq:REF_Er_IID}) and (\ref{eq:REF_E0_is}).
Using the identities $\mathbb{E}[Z]=\frac{dM_{Z}}{d\tau}\Big|_{\tau=0}$
and $\mathrm{Var}[Z]=\frac{d^{2}M_{Z}}{d\tau^{2}}\Big|_{\tau=0}$
, we obtain
\begin{align}
\mathbb{E}[Z] & =R-\frac{\partial E_{0}(Q,\rho,s)}{\partial\rho}\Big|_{\rho=\hat{\rho}}=0\label{eq:REF_Step7}\\
\mathrm{Var}[Z] & =-\frac{\partial^{2}E_{0}(Q,\rho,s)}{\partial\rho^{2}}\Big|_{\rho=\hat{\rho}}>0,\label{eq:REF_Step8}
\end{align}
where the second equality in (\ref{eq:REF_Step7}) and the inequality
in (\ref{eq:REF_Step8}) hold since $R\in\big(R_{\mathrm{cr}}(Q),I_{\mathrm{GMI}}(Q)\big)$
and hence $\hat{\rho}\in(0,1)$ (e.g. see \cite[pp. 142-143]{Gallager}).
Writing the expectation in (\ref{eq:SA_Above_3}) as a nested expectation
given $U$ and applying Lemma \ref{lem:REF_Lem20}, it follows that
\begin{align}
\overline{p}_{e} & \le K_{4}e^{-nE_{r}(Q,R)}\mathbb{E}\bigg[\frac{1}{\sqrt{n}}e^{-\hat{\rho}\log(U\sqrt{n})}\bigg]\label{eq:REF_Step9}\\
 & =K_{4}e^{-nE_{r}(Q,R)}\mathbb{E}\bigg[\frac{1}{\sqrt{n}}\bigg(\frac{1}{U\sqrt{n}}\bigg)^{\hat{\rho}}\bigg]\\
 & =\frac{K_{4}}{n^{\frac{1}{2}(1+\hat{\rho})}}e^{-nE_{r}(Q,R)}\mathbb{E}\big[U^{-\hat{\rho}}\big]\\
 & =\frac{K_{5}}{n^{\frac{1}{2}(1+\hat{\rho})}}e^{-nE_{r}(Q,R)},
\end{align}
where $K_{4}$ and $K_{5}=K_{4}\mathbb{E}\big[U^{-\hat{\rho}}\big]$
are constants. This concludes the proof.

\subsection{\label{sub:REF_PROOF_IRREGULAR}Proof for the Irregular Case}

The upper bound of one at rates below the critical rate in (\ref{eq:REF_alpha-1})
was given by Kaplan and Shamai \cite{Compound}, so we focus on rates
above the critical rate. The proof for the regular case used two applications
of Lemma \ref{lem:REF_Lem20}; see (\ref{eq:REF_Tail1}) and (\ref{eq:REF_Step9}).
The former leads to a multiplicative $n^{-\frac{\hat{\rho}(R)}{2}}$
term in the final expression, and the second leads to a multiplicative
$n^{-\frac{1}{2}}$ term. In the irregular case, we only perform the
latter application of Lemma \ref{lem:REF_Lem20}. The proof is otherwise
essentially identical. Applying Markov's inequality to the RCU bound
in (\ref{eq:REF_RCU}), we obtain 
\begin{equation}
\overline{p}_{e}\le\mathbb{E}\bigg[\min\Big\{1,Me^{-i_{s}^{n}(\boldsymbol{X},\boldsymbol{Y})}\Big\}\bigg].
\end{equation}
Repeating the analysis of the regular case starting from (\ref{eq:REF_Step3}),
we obtain the desired result.

\appendix
Here we provide the proof of Lemma \ref{lem:SetF}. The first property
is easily proved by writing
\begin{align}
 & \mathrm{Var}[i_{s}^{n}(\boldsymbol{X}_{s},\boldsymbol{Y})\,|\,\boldsymbol{Y}=\boldsymbol{y}]\\
 & \qquad=\sum_{i=1}^{n}\mathrm{Var}[i_{s}(X_{s,i},Y_{i})\,|\, Y_{i}=y_{i}]\\
 & \qquad\ge\sum_{y\in\mathcal{Y}_{1}}n\hat{P}_{\boldsymbol{y}}(y)\mathrm{Var}[i_{s}(X_{s},Y)\,|\, Y=y].
\end{align}
Substituting the bound on $\hat{P}_{\boldsymbol{y}}(y)$ in (\ref{eq:SetFn})
and the definition of $v_{s}$ in (\ref{eq:v_s}), we obtain (\ref{eq:VarLB}).
To prove that $v_{s}>0$, we note that the variance of a random variable
is zero if and only if the variable is deterministic, and hence
\begin{align}
 & \mathrm{Var}[i_{s}(X_{s},Y)\,|\, Y=y]=0\nonumber \\
 & \iff\log\frac{V_{s}(x|y)}{Q(x)}\text{ is independent of }\nonumber \\
 & \qquad\qquad\qquad\qquad\qquad\, x\text{ wherever }V_{s}(x|y)>0\\
 & \iff\frac{q(x,y)^{s}}{\sum_{\overline{x}}Q(\overline{x})q(\overline{x},y)^{s}}\text{ is independent of }\nonumber \\
 & \qquad\qquad\qquad\qquad x\text{ wherever }Q(x)q(x,y)^{s}>0\label{eq:REF_VarProof2}\\
 & \iff q(x,y)\text{ is independent of }\nonumber \\
 & \qquad\qquad\qquad\qquad\,\, x\text{ wherever }Q(x)q(x,y)>0\label{eq:REF_VarProof3}\\
 & \iff y\notin\mathcal{Y}_{1},\label{eq:REF_VarProof4}
\end{align}
where (\ref{eq:REF_VarProof2}) follows from the definition of $V_{s}$
in (\ref{eq:REF_Vs}), (\ref{eq:REF_VarProof3}) follows from the
assumption $s>0$, and (\ref{eq:REF_VarProof4}) follows from (\ref{eq:REF_Assumption1})
and the definition of $\mathcal{Y}_{1}$ in (\ref{eq:REF_SetY1}).

We now turn to the proof of the second property. Modifying the RCU
bound in (\ref{eq:REF_RCU}) to include the condition $\boldsymbol{Y}\notin\mathcal{F}_{n,\delta}$
in (\ref{eq:ErrorFn}), we have for any $s\ge0$ that
\begin{align}
 & \mathbb{P}\big[\mathrm{error}\,\cap\,\boldsymbol{Y}\notin\mathcal{F}_{n,\delta}\big]\\
 & \quad\le\sum_{\boldsymbol{x},\boldsymbol{y}\notin\mathcal{F}_{n,\delta}}Q^{n}(\boldsymbol{x})W^{n}(\boldsymbol{y}|\boldsymbol{x})\nonumber \\
 & \qquad\qquad\times\min\Big\{1,M\mathbb{P}\big[i_{s}^{n}(\overline{\boldsymbol{X}},\boldsymbol{y})\ge i_{s}^{n}(\boldsymbol{x},\boldsymbol{y})\big]\Big\}\\
 & \quad\le\sum_{\boldsymbol{x},\boldsymbol{y}\notin\mathcal{F}_{n,\delta}}Q^{n}(\boldsymbol{x})W^{n}(\boldsymbol{y}|\boldsymbol{x})\Big(Me^{-i_{s}^{n}(\boldsymbol{x},\boldsymbol{y})}\Big)^{\rho}\label{eq:REF_FnStep3}
\end{align}
where (\ref{eq:REF_FnStep3}) follows from Markov's inequality and
since $\min\{1,\alpha\}\le\alpha^{\rho}$ ($0\le\rho\le1$). We henceforth
choose $\rho$ and $s$ to achieve the maximum and supremum in (\ref{eq:REF_Er_IID})
and (\ref{eq:REF_E0_IID}) respectively, in accordance with Lemma
\ref{lem:REF_sLem}. With these choices, we have similarly to (\ref{eq:REF_E0_is})
that
\begin{equation}
e^{-nE_{r}(Q,R)}=\sum_{\boldsymbol{x},\boldsymbol{y}}Q^{n}(\boldsymbol{x})W^{n}(\boldsymbol{y}|\boldsymbol{x})\Big(Me^{-i_{s}^{n}(\boldsymbol{x},\boldsymbol{y})}\Big)^{\rho}.
\end{equation}
Hence, we will complete the proof by showing that
\begin{equation}
\sum_{\boldsymbol{x},\boldsymbol{y}\notin\mathcal{F}_{n,\delta}}Q^{n}(\boldsymbol{x})W^{n}(\boldsymbol{y}|\boldsymbol{x})e^{-\rho i_{s}^{n}(\boldsymbol{x},\boldsymbol{y})}\label{eq:REF_Quantity1}
\end{equation}
has a strictly larger exponential rate of decay than
\begin{equation}
\sum_{\boldsymbol{x},\boldsymbol{y}}Q^{n}(\boldsymbol{x})W^{n}(\boldsymbol{y}|\boldsymbol{x})e^{-\rho i_{s}^{n}(\boldsymbol{x},\boldsymbol{y})}\label{eq:REF_Quantity2}
\end{equation}
for some $\delta>0$. By performing an expansion in terms of types,
(\ref{eq:REF_Quantity2}) is equal to 
\begin{align}
 & \sum_{P_{XY}\in\mathcal{P}_{n}(\mathcal{X}\times\mathcal{Y})}\mathbb{P}\big[(\boldsymbol{X},\boldsymbol{Y})\in T^{n}(P_{XY})\big]e^{-n\rho\mathbb{E}_{P}[i_{s}(X,Y)]}\\
 & =\max_{P_{XY}}\exp\Big(-n\big(D(P_{XY}\|Q\times W)\nonumber \\
 & \qquad\qquad\qquad\qquad+\rho\mathbb{E}_{P}[i_{s}(X,Y)]+o(1)\big)\Big),\label{eq:REF_FnStep6}
\end{align}
where (\ref{eq:REF_FnStep6}) follows from the property of types in
\cite[Eq. (12)]{GallagerCC} and the fact that the number of joint
types is polynomial in $n$. Substituting the definitions of divergence
and $i_{s}$ (see (\ref{eq:REF_is})) into (\ref{eq:REF_FnStep6}),
we see that the exponent of (\ref{eq:REF_Quantity2}) equals 
\begin{multline}
\min_{P_{XY}}\sum_{x,y}P_{XY}(x,y)\\
\times\log\Bigg(\frac{P_{XY}(x,y)}{Q(x)W(y|x)}\Bigg(\frac{q(x,y)^{s}}{\sum_{\overline{x}}Q(\overline{x})q(\overline{x},y)^{s}}\Bigg)^{\rho}\Bigg).\label{eq:LessConstr}
\end{multline}
Similarly, and from the definition of $\mathcal{F}_{n,\delta}$ in
(\ref{eq:SetFn}), (\ref{eq:REF_Quantity1}) has an exponent equal
to 
\begin{multline}
\min_{P_{XY}\,:\,\sum_{y\in\mathcal{Y}_{1}}P_{Y}(y)\le\delta}\sum_{x,y}P_{XY}(x,y)\\
\times\log\Bigg(\frac{P_{XY}(x,y)}{Q(x)W(y|x)}\Bigg(\frac{q(x,y)^{s}}{\sum_{\overline{x}}Q(\overline{x})q(\overline{x},y)^{s}}\Bigg)^{\rho}\Bigg).\label{eq:MoreConstr}
\end{multline}
A straightforward evaluation of the Karush-Kuhn-Tucker (KKT) conditions
\cite[Sec. 5.5.3]{Convex} yields that (\ref{eq:LessConstr}) is uniquely
minimized by 
\begin{multline}
P_{XY}^{*}(x,y)\\
=\frac{Q(x)W(y|x)\bigg(\frac{\sum_{\overline{x}}Q(\overline{x})q(\overline{x},y)^{s}}{q(x,y)^{s}}\bigg)^{\rho}}{\sum_{x^{\prime},y^{\prime}}Q(x^{\prime})W(y^{\prime}|x^{\prime})\bigg(\frac{\sum_{\overline{x}^{\prime}}Q(\overline{x}^{\prime})q(\overline{x}^{\prime},y^{\prime})^{s}}{q(x^{\prime},y^{\prime})^{s}}\bigg)^{\rho}}.\label{eq:P*XY}
\end{multline}
From the assumptions in (\ref{eq:REF_Assumption1}) and (\ref{eq:REF_Assumption2}),
we can find a symbol $y^{*}\in\mathcal{Y}_{1}$ such that $P_{Y}^{*}(y^{*})>0$.
Choosing $\delta<P_{Y}^{*}(y^{*})$, it follows that $P_{XY}^{*}$
fails to satisfy the constraint in (\ref{eq:MoreConstr}), and thus
(\ref{eq:MoreConstr}) is strictly greater than (\ref{eq:LessConstr}).

 \bibliographystyle{IEEEtran}
\bibliography{12-Paper,18-MultiUser,18-SingleUser}
 
\end{document}